\documentclass[final,5p,times,twocolumn]{elsarticle}
\usepackage{lineno,hyperref,amsmath,amsthm,amssymb,amsfonts,ragged2e,color,subfig}
\journal{``Communications in Theoretical Physics"}
\begin{document}
\begin{frontmatter}
\title{Dust-acoustic rogue waves in an electron-positron-ion-dust plasma medium}
\author{M. H. Rahman$^{*,1}$, N. A. Chowdhury$^{\dag,1,2}$, A. Mannan$^{\ddag,1,3}$, and A. A. Mamun$^{\S,1,4}$}
\address{$^{1}$Department of Physics, Jahangirnagar University, Savar, Dhaka-1342, Bangladesh\\
$^{2}$Plasma Physics Division, Atomic Energy Centre, Dhaka-1000, Bangladesh\\
$^{3}$Institut f\"{u}r Mathematik, Martin Luther Universit\"{a}t Halle-Wittenberg, Halle, Germany\\
$^{4}$Wazed Miah Science Research Center, Jahangirnagar University, Savar, Dhaka-1342, Bangladesh\\
e-mail: $^*$rahman1992phy@gmail.com, $^{\dag}$nurealam1743phy@gmail.com,\\ $^\ddag$abdulmannan@juniv.edu, $^\S$mamun\_phys@juniv.edu}
\begin{abstract}
A precise theoretical investigation has been made on dust-acoustic (DA) waves (DAWs) in a
  four components dusty plasma medium having inertial warm adiabatic dust grains and inertialess
  $q$-distributed electrons as well as isothermal ions and positrons. The nonlinear and dispersive parameters
  of the nonlinear Schr\"{o}dinger equation (NLSE), which develops by using reductive perturbations technique,
  have been used to recognize the stable and unstable parametric regions of the DAWs as well as associated DA
  rogue waves (DARWs) in the unstable parametric regime of the DAWs. The effects of the light positrons and massive
  dust grains in determining the amplitude and width of the DARWs associated with DAWs are examined. The findings
  of our present investigation may be useful for understanding different nonlinear electrostatic phenomena
  in both space and laboratory plasmas.
\end{abstract}
\end{frontmatter}
\section{Introduction}
\label{3sec:Introduction}
The electron-positron-ion-dust (EPID) plasma is a fully ionized gas
comprises of electrons and positrons of equal masses but opposite polarity as well as
ions and massive dust grains \cite{C1,C2,Esfandyari-Kalejahi2012,Jehan2009}. The simultaneous co-existence of the light positrons and
heavy dust grains as well as electrons and ions in contrast to the usual plasma
containing of electrons and ions in space and laboratory EPID plasma medium (EPIDPM) has rigourously
changed the basic features of the propagation of nonlinear electrostatic waves, viz.,
Dust-acoustic (DA) waves (DAWs) \cite{Esfandyari-Kalejahi2012,Jehan2009,Emamuddin2013,Eslami2011},
DA solitary waves (DA-SWs) \cite{Esfandyari-Kalejahi2012,Jehan2009}, DA shock waves (DA-SHWs) \cite{Roy2014},
DA rogue waves (DA-RWs) \cite{Rahman2018a,Bains2013,Moslem2011,Rahman2018b}, dust-ion-acoustic
waves (DI-AWs) \cite{Banerjee2016,Paul2016} in an EPIDPM, and
has received much attraction to solve the profound mystery, and  has identified in space plasmas, viz., active
galactic nuclei \cite{Esfandyari-Kalejahi2012,Jehan2009}, pulsar magnetospheres  \cite{Esfandyari-Kalejahi2012,Jehan2009}, interstellar clouds \cite{Esfandyari-Kalejahi2012,Jehan2009}, supernova environments \cite{Esfandyari-Kalejahi2012,Jehan2009},
early universe \cite{Banerjee2016}, the inner regions of the accretion disks surrounding the black hole
as well as in laboratory experiments of cluster explosions by intense laser beams \cite{Higdon2009}.

A new velocity distribution recognized as ``Tsallis statistics" has attracted much
interest among the plasma physicists and is believed to be a useful generalization of the
conventional Boltzmann-Gibbs statistics which is not valid for describing the long-range
interaction in system such as complex plasma system, and also has considered as an inevitable
tool for analysing and predicting the statistical features of long-range interaction in complex system.
This Tsallis distribution was first proposed by Tsallis \cite{Tsallis1988} and was first
acknowledged by Renyi \cite{Renyi1955}. The degree of non-extensivity of the long-range interaction in complex system
is denoted by the entropic index $q$, and when $(q \rightarrow1)$, Tsallis distribution reduces to
well known Maxwell-Boltzmann velocity distribution \cite{Emamuddin2013,Eslami2011,Roy2014}.
Emamuddin \textit{et al.} \cite{Emamuddin2013} examined DA Gardner solitons in a four components
dusty plasma medium (DPM) having non-extensive plasma species.  Eslami \textit{et al.} \cite{Eslami2011}  analyzed
the properties of DA-SWs in presence of non-extensive plasma species, and found that the DA-SWs
exhibit compression and rarefaction according to the negative and positive $q$, respectively.
Roy \textit{et al.} \cite{Roy2014} demonstrated DA-SHWs in a three components DPM featuring non-extensive
electrons, and observed that DA-SHWs potential decreases with the increase in the value of $q$ for positive limit.

The signature of the positrons in EPIDPM has encouraged many authors
to examine the propagation of nonlinear electrostatic pulses in EPIDPM because the effect
of positrons in EPIDPM with substantial amount can not be ignored. Banerjee and Maitra \cite{Banerjee2016}
considered four components EPIDPM and studied electrostatic potential structures in presence of the massive dust grains and light positrons
and observed that the hight of the potential structures increases with increasing value of massive dust number density but decreases with
increasing light positron number density. Paul and Bandyopadhyay \cite{Paul2016} demonstrated DI-AWs in an EPIDPM and highlighted the existence of
both polarities solitary waves as well as positive and negative potential double layers but only positive super-solitons in an EPIDPM having
positrons.

The nonlinear Schr\"{o}dinger equation (NLSE) \cite{Rahman2018a,Bains2013,Moslem2011,Rahman2018b,C3,C4,C5} has considered
one of the elegant equations which can art the picture of the modulational instability (MI) and the mechanism
of rogue waves (RWs) in any complex medium, and also has manifested as a central archetype to investigate
the nonlinear property of complex plasma medium. Bains \textit{et al.} \cite{Bains2013} studied the propagation
of DAWs in a three components DPM having non-extensive plasma species and found that the critical wave number ($k_c$) increases
with increasing the non-extensivity of the plasma species. Moslem \textit{et al.} \cite{Moslem2011} analyzed the DAWs
by considering $q$-distributed plasma species and observed that non-extensive plasma parameter $q$ plays a
significant role in maximizing or minimizing the energy of the DA-RWs. Rahman \textit{et al.} \cite{Rahman2018b}
examined DA-RWs in a multi-component DPM and observed that the ion temperature enhances the height and thickness of the
DA-RWs while the electron temperature reduces the height and thickness of the DA-RWs.

Recently, Esfandyari-Kalejahi \textit{et al.} \cite{Esfandyari-Kalejahi2012} investigated the nonlinear propagation of
DA-SWs in an EPIDPM and observed that the amplitude of DA-SWs increases with the increase in the value of
negatively charged dust charge state. Jehan \textit{et al.} \cite{Jehan2009}
studied DA-SWs in a four components DPM by considering inertialess iso-thermal electrons, positrons, and ions as well as
inertial massive dust grains, and found that the amplitude of dip solitons increases with increasing the value of the
positron concentration. However, to the best knowledge of the authors, there is no theoretical investigation has been made on DAWs in
EPIDPM considering warm adiabatic dust grains. Therefore, in this paper, we are interested to study
the MI of DAWs and the formation of DA-RWs in an EPIDPM as well as the effects of $q$-distributed electrons, and isothermal
ions and positrons on the DA-RWs in a four components EPIDPM.

The rest of the paper is organized in the following fashion: The governing equations
of our plasma model are presented in Sec. \ref{3sec:Governing equations}. The MI and 
RWs are given in Sec. \ref{3sec:Modulational instability and rogue waves}.
Finally, a brief conclusion is provided in Sec. \ref{3sec:Conclusion}.
\section{Governing equations}
\label{3sec:Governing equations}
We consider the propagation of DAWs in an unmagnetized collisionless EPIDPM consisting of
inertial warm adiabatic negatively charged massive dust grains (mass $m_d$; charge $q_d=-Z_de$), and inertialess
$q$-distributed electrons (mass $m_e$; charge $-e$) as well as iso-thermal ions (mass $m_i$; charge $q_i=+Ze$)
and positrons (mass $m_p$; charge $+e$), where $Z_d$ ($Z_i$) is the number of electron (proton) residing on a 
negatively (positively) charged massive dust grains (ions).
Overall, the charge neutrality condition for our plasma model can be written as
$n_{e0} +Z_{d} n_{d0}= n_{p0}+Z_in_{i0}$. Now, the basic set of normalized equations can be written in the form
\begin{eqnarray}
&&\hspace*{-1.3cm}\frac{\partial n_d}{\partial t}+\frac{\partial(n_d u_d)}{\partial x}=0,
\label{3:eq1}\\
&&\hspace*{-1.3cm}\frac{\partial u_d}{\partial t} + u_d\frac{\partial u_d}{\partial x}+\delta n_d\frac{\partial n_d}{\partial x}=\frac{\partial \varphi}{\partial x},
\label{3:eq2}\\
&&\hspace*{-1.3cm}\frac{\partial^2 \varphi}{\partial x^2}=(\mu_p+\mu_i-1) n_e-\mu_p n_p-\mu_i n_i+n_d,
\label{3:eq3}
\end{eqnarray}
where $n_d$ is the adiabatic dust grains number density normalized by its equilibrium value $n_{d0}$;
$u_d$ is the dust fluid speed normalized by the DA wave speed $C_d=(Z_dK_BT_i/m_d)^{1/2}$
(with $T_i$ being the ion temperature, $m_d$ being the dust grain mass, and $K_B$ being the Boltzmann
constant); $\phi$ is the electrostatic wave potential normalized by $K_BT_i/e$ (with $e$ being the
magnitude of single electron charge); the time and space variables are normalized by
$\omega_{pd}^{-1}=(m_d/4\pi Z_d^2e^2n_{d0})^{1/2}$ and $\lambda_{Dd}=(K_BT_i/4\pi Z_dn_{d0}e^2)^{1/2}$,
respectively; $P_d=P_{d0}(N_d/n_{d0})^\gamma$ [with $P_{d0}$ being the equilibrium adiabatic
pressure of the dust, and $\gamma=(N+2)/N$, where $N$ is the degree of freedom and for one-dimensional case,
$N=1$ then $\gamma=3$]; $P_{d0}=n_{d0}K_BT_d$ (with $T_d$ being the temperatures of the adiabatic dust grains);
and other plasma parameters are considered as $\delta=3T_d/Z_dT_i$, $\mu_p=n_{p0}/Z_dn_{d0}$, and $\mu_i=Z_in_{i0}/Z_dn_{d0}$.
Now, the expression for electron number density obeying $q$-distribution is given by \cite{Tsallis1988,Rahman2018a}
\begin{eqnarray}
&&\hspace*{-1.3cm}n_e= \left[1+(q-1)\sigma\varphi\right]^{\frac{q+1}{2(q-1)}},
\label{3:eq4}
\end{eqnarray}
where $\sigma=T_i/T_e$ and $T_e>T_i$. The expression for the number density of ions and positrons
obeying iso-thermal distribution  are, respectively, given by
\begin{eqnarray}
&&\hspace*{-1.3cm}n_p=\mbox{exp}(-\alpha \varphi),
\label{3:eq5}\\
&&\hspace*{-1.3cm}n_i=\mbox{exp}(-\varphi),
\label{3:eq6}
\end{eqnarray}
where $\alpha=T_i /T_p$ and $T_p>T_i$. Now, by substituting Eqs. \eqref{3:eq4}-\eqref{3:eq6} in Eq. \eqref{3:eq3},
and expanding up to third order in $\varphi$, we get
\begin{eqnarray}
&&\hspace*{-1.3cm}\frac{\partial^2 \varphi}{\partial x^2}+1=n_d+\Upsilon_1 \varphi+\Upsilon_2\varphi^2+\Upsilon_3 \varphi^3+\cdot\cdot\cdot,
\label{3:eq7}
\end{eqnarray}
where
\begin{eqnarray}
&&\hspace*{-1.3cm}\Upsilon_1=[\sigma(q+1)(\mu_p+\mu_i-1)+2\mu_p\alpha+2\mu_i]/2,
\nonumber\\
&&\hspace*{-1.3cm}\Upsilon_2=[\sigma^2(q+1)(q-3)(1-\mu_p-\mu_i)-4\mu_p\alpha^2-4\mu_{i}]/8,
\nonumber\\
&&\hspace*{-1.3cm}\Upsilon_3=[\sigma^3(q+1)(q-3)(3q-5)(\mu_p+\mu_i-1)+F1]/48,
\nonumber\
\end{eqnarray}
where $F1=8\mu_p\alpha^3+8\mu_{i}$. To study the MI of the DAWs, we want to
derive the NLSE by employing the reductive perturbation method and for that case, first we can write
the stretched coordinates as $\xi={\varepsilon}(x-v_gt)$ and $\tau={\varepsilon}^2 t$,
where $\varepsilon$ is a smallness parameter and $v_g$ is the group velocity of DAWs
to be determined later. Now, we can write the dependent variables as \cite{C6,C7,C8}
\begin{eqnarray}
&&\hspace*{-1.3cm}n_{d}=1 +\sum_{m=1}^{\infty}\epsilon^{m}\sum_{l=-\infty}^{\infty}n_{dl}^{(m)}(\xi,\tau)~\mbox{exp}[i l(kx-\omega t)],
\label{3:eq8}\\
&&\hspace*{-1.3cm}u_{d}=\sum_{m=1}^{\infty}\epsilon^{m}\sum_{l=-\infty}^{\infty}u_{dl}^{(m)}(\xi,\tau)~\mbox{exp}[i l(kx-\omega t)],
\label{3:eq9}\\
&&\hspace*{-1.3cm}\varphi=\sum_{m=1}^{\infty}\epsilon^{m}\sum_{l=-\infty}^{\infty}\varphi_{l}^{(m)}(\xi,\tau)~\mbox{exp}[i l(kx-\omega t)],
\label{3:eq10}
\end{eqnarray}
where $k$ and $\omega$ is real variables presenting the
carrier wave number and frequency, respectively. The derivative operators can be written as
\begin{eqnarray}
&&\hspace*{-1.3cm}\frac{\partial}{\partial t}\rightarrow\frac{\partial}{\partial t}-\epsilon v_g \frac{\partial}{\partial\xi}+\epsilon^2\frac{\partial}{\partial\tau},
\label{3:eq11}\\
&&\hspace*{-1.3cm}\frac{\partial}{\partial x}\rightarrow\frac{\partial}{\partial x}+\epsilon\frac{\partial}{\partial\xi}.
\label{3:eq12}
\end{eqnarray}
Now, by substituting Eqs. \eqref{3:eq8}-\eqref{3:eq12} into Eqs. \eqref{3:eq1}, \eqref{3:eq2}, and \eqref{3:eq7}, and
taking the terms containing $\varepsilon$, the first-order ($m=1$ with $l=1$) reduced equation can be written as
\begin{eqnarray}
&&\hspace*{-1.3cm}n_{d1}^{(1)}=\frac{k^2}{\Omega}\varphi_1^{(1)},~~~~~~~ u_{d1}^{(1)}=\frac{k\omega}{\Omega}\varphi_1^{(1)},
\label{3:eq13}
\end{eqnarray}
where $\Omega=\delta k^2-\omega^2$. We thus obtain the dispersion relation for DAWs
\begin{eqnarray}
&&\hspace*{-1.3cm}\omega^2=\frac{k^2}{\Upsilon_{1}+k^2}+\delta k^2,
\label{3:eq14}
\end{eqnarray}
The second-order ($m=2$ with $l=1$) equations are given by
\begin{eqnarray}
&&\hspace*{-1.3cm}n_{d1}^{(2)}=\frac{k^2}{\Omega}\varphi_1^{(2)}-\frac{2ik\omega(v_{g}k-\omega)}{\Omega^2}\frac{\partial \varphi_1^{(1)}}{\partial\xi},
\label{3:eq15}\\
&&\hspace*{-1.3cm}u_{d1}^{(2)}=\frac{k\omega}{\Omega}\varphi_1^{(2)}-\frac{i(v_{g}k-\omega)(\omega^2+\delta k^2)}{\Omega^2}\frac{\partial\varphi_1^{(1)}}{\partial\xi},
\label{3:eq16}
\end{eqnarray}
with the compatibility condition
\begin{eqnarray}
&&\hspace*{-1.3cm}v_g=\frac{\partial \omega}{\partial k}=\frac{\omega^2-\Omega^2}{k \omega}.
\label{3:eq17}
\end{eqnarray}
The coefficients of $\epsilon$ for $m=2$ and $l=2$ provide the second order harmonic
amplitudes which are found to be proportional to $|\varphi_1^{(1)}|^2$
\begin{eqnarray}
&&\hspace*{-1.3cm}n_{d2}^{(2)}=B_1|\varphi_1^{(1)}|^2,
\label{3:eq18}\\
&&\hspace*{-1.3cm}u_{d2}^{(2)}=B_2 |\varphi_1^{(1)}|^2,
\label{3:eq19}\\
&&\hspace*{-1.3cm}\varphi_2^{(2)}=B_3 |\varphi_1^{(1)}|^2,
\label{3:eq20}
\end{eqnarray}
where
\begin{eqnarray}
&&\hspace*{-1.3cm}B_1=\frac{2B_3 k^2 \Omega^2  -3 \omega^2 k^4-\delta k^6}{2\Omega^3},
\nonumber\\
&&\hspace*{-1.3cm}B_2=\frac{2 B_1\delta k \Omega^2 +\delta k^5+\omega^2k^3-2B_3k\Omega^2}{2\omega \Omega^2},
\nonumber\\
&&\hspace*{-1.3cm}B_3=\frac{3\omega^2 k^4 +\delta k^6-2\Upsilon_2\Omega^3}{6k^2\Omega^3}.
\nonumber\
\end{eqnarray}
Now, we consider the expression for ($m=3$ with $l=0$) and ($m=2$ with $l=0$),
which leads the zeroth harmonic modes. Thus, we obtain
\begin{eqnarray}
&&\hspace*{-1.3cm}n_{d0}^{(2)}=B_4 |\varphi_1^{(1)}|^2,
\label{3:eq21}\\
&&\hspace*{-1.3cm}u_{d0}^{(2)}=B_5|\varphi_1^{(1)}|^2,
\label{3:eq22}\\
&&\hspace*{-1.3cm}\varphi_0^{(2)}=B_6 |\varphi_1^{(1)}|^2,
\label{3:eq23}
\end{eqnarray}
where
\begin{eqnarray}
&&\hspace*{-1.3cm}B_4=\frac{ B_6 \Omega^2-2v_g\omega k^3-\delta k^4-k^2\omega^2 }{\Omega^2(\delta-v_g^2)},
\nonumber\\
&&\hspace*{-1.3cm}B_5=\frac{B_4\delta \Omega^2-B_6\Omega^2+\delta k^4+k^2\omega^2}{v_g\Omega^2},
\nonumber\\
&&\hspace*{-1.3cm}B_6=\frac{2 v_g \omega k^3+\delta k^4+k^2\omega^2-2\Upsilon_2\Omega^2(\delta-v_g^2)}{\Omega^2(1+\Upsilon_1\delta-\Upsilon_1v^2_g)}.
\nonumber\
\end{eqnarray}
Finally, the third harmonic modes ($m=3$) and ($l=1$), with the help of \eqref{3:eq13}$-$\eqref{3:eq23},
give a set of equations, which can be reduced to the following  NLSE:
\begin{eqnarray}
&&\hspace*{-1.3cm}i\frac{\partial \phi}{\partial \tau}+P\frac{\partial^2 \phi}{\partial \xi^2}+Q|\phi|^2\phi=0,
\label{3eq:24}
\end{eqnarray}
where $\phi=\varphi_1^{(1)}$ for simplicity. In Eq. \eqref{3eq:24}, $P$ is the dispersion coefficient
which can be written as
\begin{eqnarray}
&&\hspace*{-1.3cm}P =\frac{(v_gk-\omega)(\omega^3-3v_gk\omega^2+3\delta \omega k^2-v_g\delta k^3)-\Omega^3}{2 \omega \Omega k^2},
\nonumber\
\end{eqnarray}
and $Q$ is the nonlinear coefficient
which can be written as
\begin{eqnarray}
&&\hspace*{-1.3cm}Q=\frac{3\Upsilon_3 \Omega^2+2\Upsilon_2\Omega^2(B_3+B_6)-2\omega k^3(B_2+B_5)-F2}{2\omega k^2},
\nonumber\
\end{eqnarray}
where $F2=(\delta k^4+k^2\omega^2)(B_1+B_4)$. The space and time evolution of the DAWs in EPIDPM are directly governed by the  dispersion
($P$) and nonlinear  ($Q$) coefficients, and indirectly governed by different plasma parameters such as
as $\delta$, $\mu_p$,  $\mu_i$, $\sigma$, $\alpha$, $q$, and $k$. Thus, these plasma parameters significantly
affect the stability conditions of the DAWs.
\begin{figure}[t!]
\includegraphics[width=80mm]{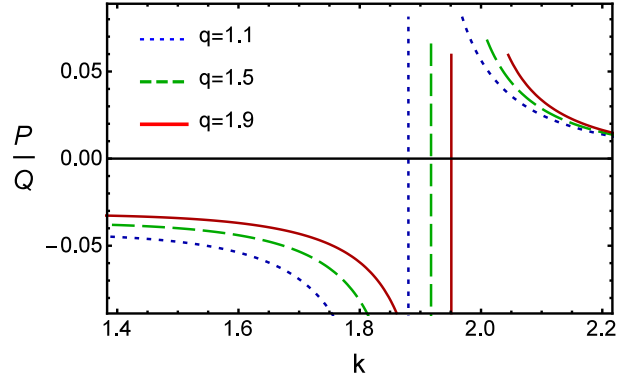}
\caption{$P/Q$ versus $k$ curve for different values of positive $q$ along with $\alpha=0.3$, $\delta=0.005$, $\mu_i=1.4$,
$\mu_p=0.3$, and $\sigma=0.7$.}
\label{3Fig:F1}
\end{figure}
\begin{figure}[t!]
\includegraphics[width=80mm]{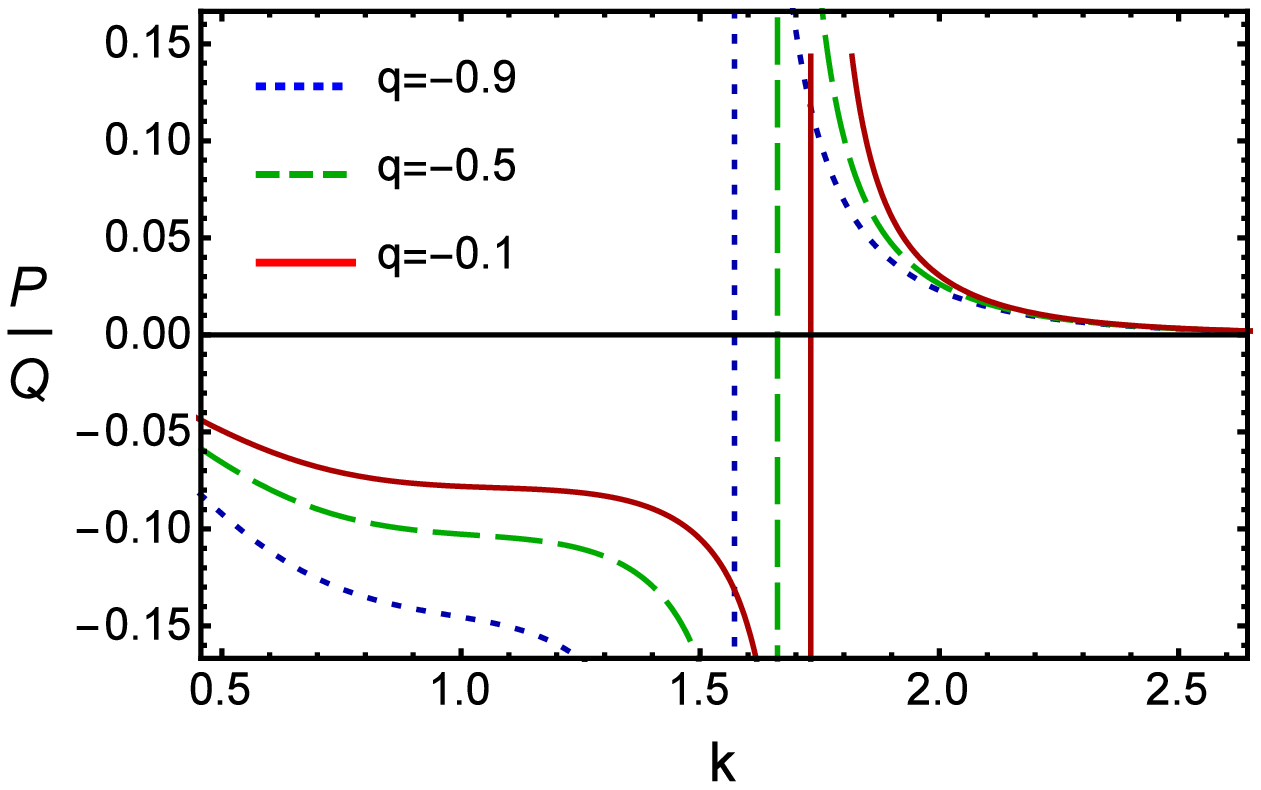}
\caption{$P/Q$ versus $k$ curve for different values of negative $q$ along with $\alpha=0.3$, $\delta=0.005$, $\mu_i=1.4$,
$\mu_p=0.3$, and $\sigma=0.7$.}
\label{3Fig:F2}
\end{figure}
\begin{figure}[t!]
\includegraphics[width=80mm]{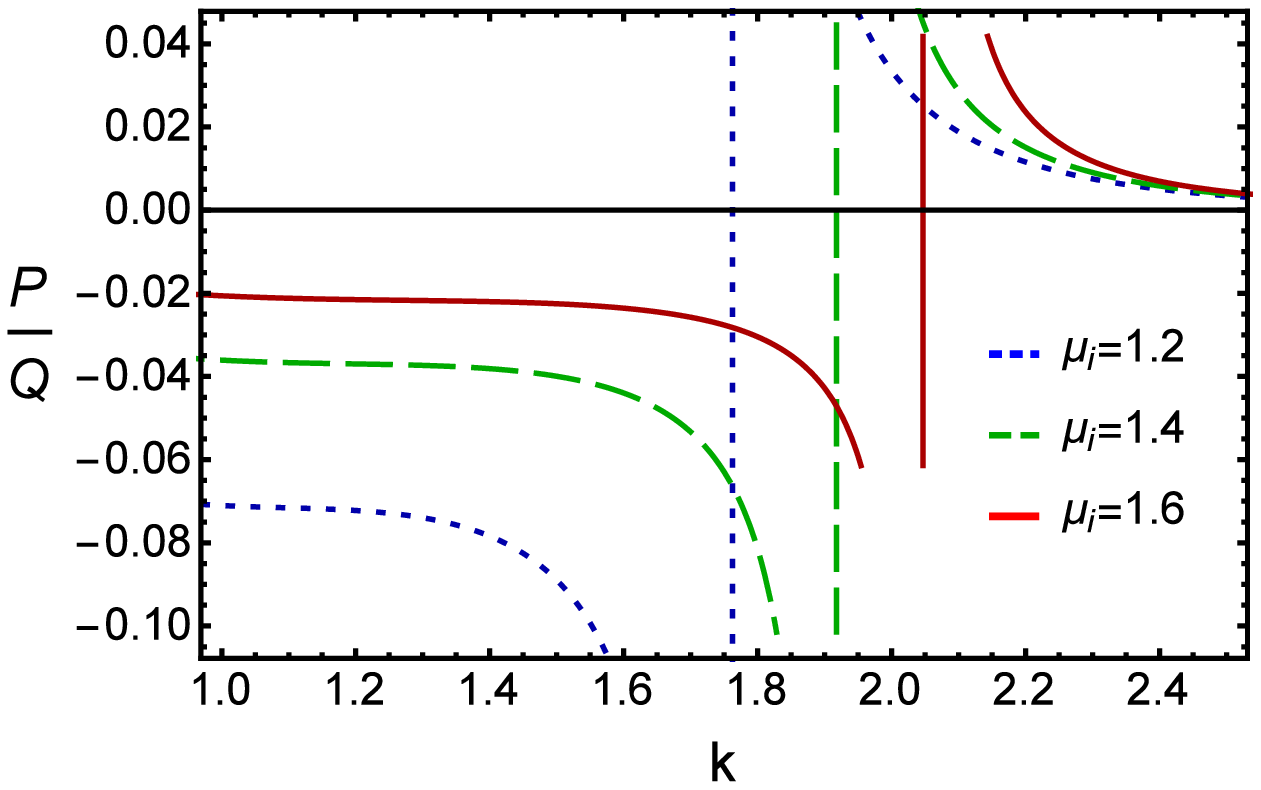}
\caption{$P/Q$ versus $k$ curve for different values of $\mu_i$ along with $\alpha=0.3$, $\delta=0.005$,
$\mu_p=0.3$, $\sigma=0.7$ and $q=1.5$.}
\label{3Fig:F3}
\end{figure}
\begin{figure}[t!]
\includegraphics[width=80mm]{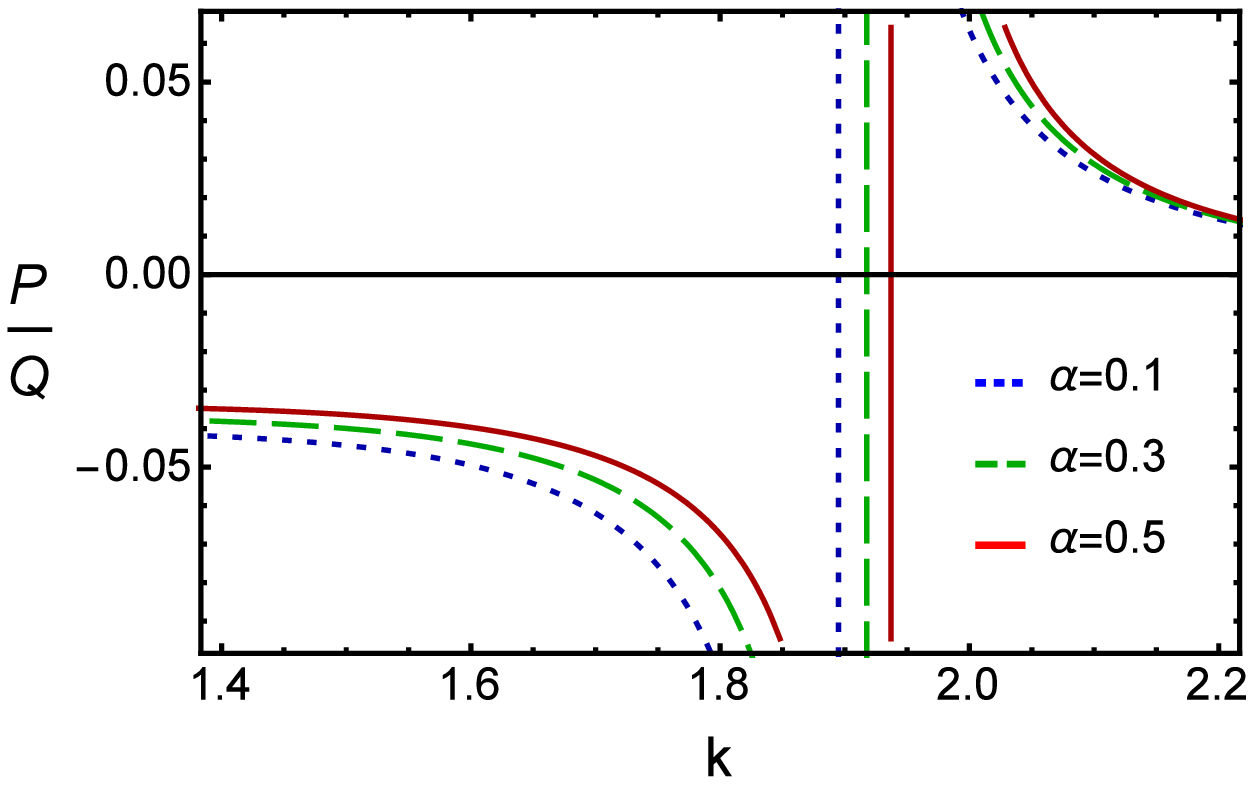}
\caption{$P/Q$ versus $k$ curve for different values of $\alpha$ along with $\delta=0.005$, $\mu_i=1.4$,
$\mu_p=0.3$, $\sigma=0.7$, and $q=1.5$.}
\label{3Fig:F4}
\end{figure}
\begin{figure}[t!]
\includegraphics[width=80mm]{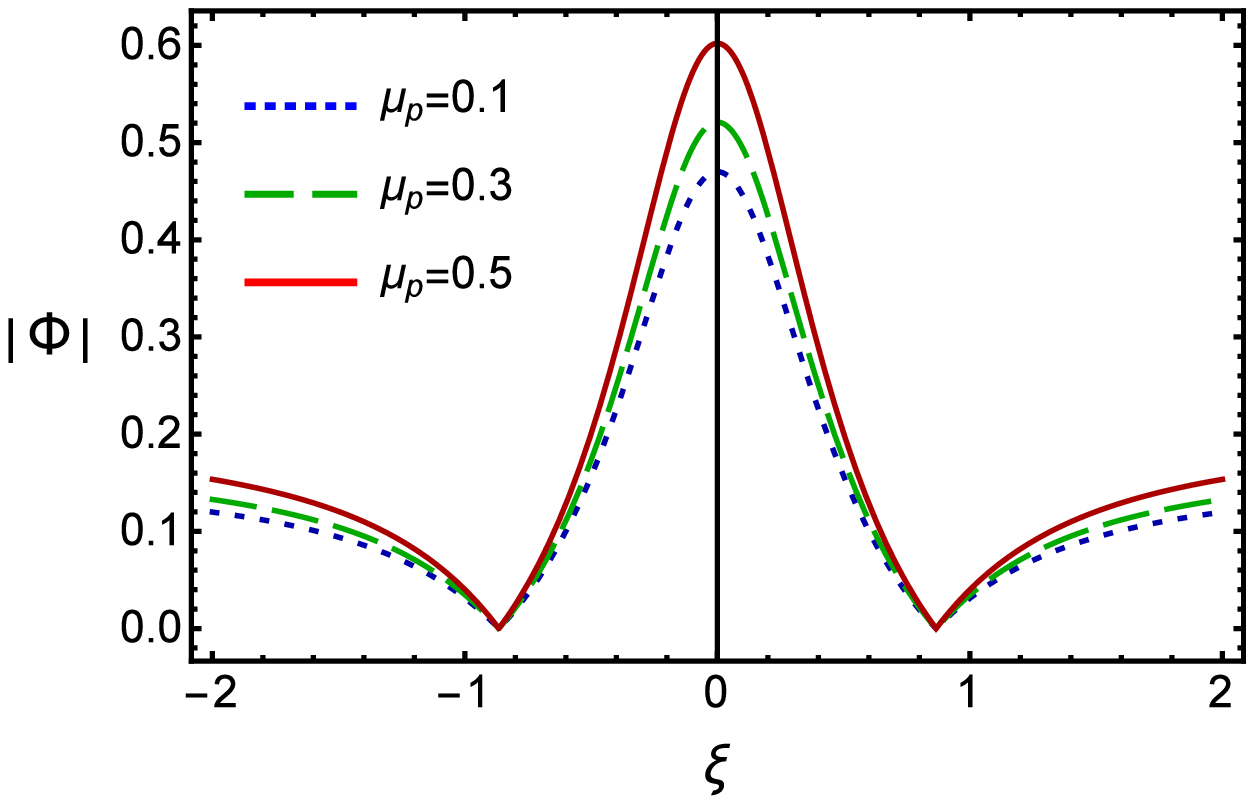}
\caption{variation of $|\phi|$ with $\xi$ for different values of $\mu_p$ along with $\alpha=0.3$, $\delta=0.005$, $\mu_i=1.4$, $\sigma=0.7$, and $q=1.5$.}
\label{3Fig:F5}
\end{figure}
\begin{figure}[t!]
\includegraphics[width=80mm]{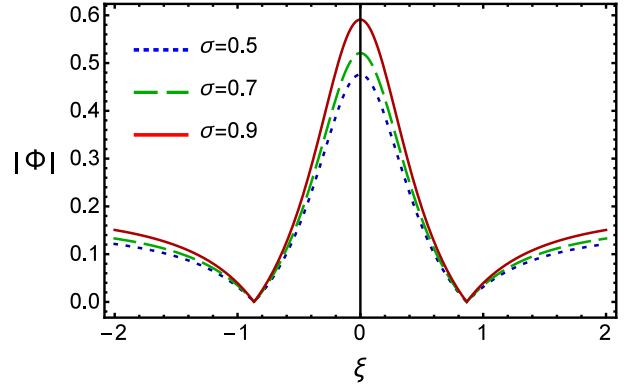}
\caption{variation of $|\phi|$ with $\xi$ for different values of $\sigma$ along with $\alpha=0.3$, $\delta=0.005$, $\mu_i=1.4$,
$\mu_p=0.3$, and $q=1.5$.}
\label{3Fig:F6}
\end{figure}
\section{Modulational instability and rogue waves}
\label{3sec:Modulational instability and rogue waves}
The stability of DAWs in an EPIDPM is governed by the sign of dispersion coefficient ($P$)
and nonlinear coefficient ($Q$) which are function of various plasma parameters such as
$\alpha$, $\delta$, $\mu_i$, $\mu_p$, $\sigma$, and $q$. These plasma parameters significantly
control the stability conditions of the DAWs. When $P$ and $Q$ are same sign ($P/Q>0$), the
evolution of the DAWs amplitude is modulationally unstable and in this region electrostatic
bright envelope solitons as well as RWs exist \cite{Rahman2018a,Fedele2002}. On the other hand, when $P$ and $Q$
are opposite sign ($P/Q<0$), the DAWs are modulationally stable in presence of external perturbations
and in this region electrostatic dark envelope solitons exist \cite{Rahman2018a,Fedele2002}. The plot of $P/Q$ against $k$ yields stable and
unstable regions of the DAWs. The point, at which transition of $P/Q$ curve intersects with $k$-axis,
is known as the threshold or critical wave number $k$ ($=k_c$).

We have depicted $P/Q$ versus $k$ graph for different values of positive $q$ and negative $q$
in Figs. \ref{3Fig:F1} and \ref{3Fig:F2}, respectively, and it is obvious from these figures that
(a) the DAWs remain stable for small $k$ ($k<k_c$) and the MI sets in for values of $k$ ($k>k_c$); (b) when $q=1.1$, $1.5$, and $1.9$
then the corresponding $k_c$ value is $k_c\equiv 1.88$ (dotted blue curve), $k_c\equiv 1.92$ (dashed green curve), and $k_c\equiv1.95$
(solid red curve); (c) $k_c$ increases with increasing value of positive $q$. On the other hand, in the case of negative $q$:
(d) when $q=-0.9$, $-0.5$, and $-0.1$ then the corresponding $k_c$ value is $k_c\equiv 1.57$ (dotted blue curve), $k_c\equiv 1.67$
(dashed green curve), and $k_c\equiv1.72$ (solid red curve); (e) the negative $q$ also increases the value of critical wave number.
Hence one can say that the stability of the DAWs is independent on the sign of $q$ but dependent on the magnitude of $q$.

We have investigated the effects of ion and dust number density as well as their charge state
to organize the stable and unstable parametric regions of DAWs by depicting $P/Q$ versus $k$ graph
for different values of  $\mu_i$ in Fig. \ref{3Fig:F3}. It can be seen from this figure that (a) the $k_c$ increases with $\mu_i$;
the stable (unstable) parametric region increases (decreases) with increasing value of ion (dust) number
density for the fixed values of $Z_d$ and $Z_i$; (b)  the $k_c$ increases with the increase in the value
of positive ion charge state ($Z_i$) while decreases with the increase in the value of negative dust
charge state ($Z_d$) for a constant value of $n_{i0}$ and $n_{d0}$ (via $\mu_i=Z_in_{i0}/Z_d n_{d0}$).
So, the charge and the number density of the positive ion and negative dust grains play an opposite role in recognizing the
stable and unstable parametric regions of the DAWs.

Figure \ref{3Fig:F4} represents the stability nature of the DAWs according to the ion to positron
temperature (via $\alpha=T_i/T_p$). The unstable window increases with an increase in ion temperature
while decreases with an increase in the positron temperature.

The governing equation for the highly energetic DA-RWs in the unstable region ($P/Q>0$)
can be written as \cite{Akhmediev2009,Ankiewiez2009}
\begin{eqnarray}
&&\hspace*{-1.3cm}\phi(\xi,\tau)=\sqrt{\frac{2P}{Q}} \left[\frac{4(1+4iP\tau)}{1+16P^2\tau^2+4\xi^2}-1 \right]\mbox{exp}(2iP\tau).
\label{3eq:25}
\end{eqnarray}
We have also numerically analyzed Eq. \eqref{3eq:25} in Fig. \ref{3Fig:F5} to illustrate the influence of number
density ratio of positron to dust (via $\mu_p=n_{p0}/Z_d n_{d0}$) on the formation of DA-RWs associated with
unstable parametric region (i.e., $P/Q>0$), and it is clear from this figure that (a) the nonlinearity as
well as the amplitude and width of the DA-RWs increases with increasing  positron number density $n_{p0}$ for
a constant value of $Z_d$ and $n_{d0}$; (b) On the other hand, the nonlinearity as well as the amplitude
and width of the DA-RWs decreases with increasing  negative dust number density $n_{d0}$ as well as negative
dust charge state $Z_d$ for a constant value of $n_{p0}$.

The effect of ion temperature ($T_i$) as well as electron temperature ($T_e$) on the height and thickness
of DA-RWs can be observed from Fig. \ref{3Fig:F6} and it can be manifested from this figure that the height
and thickness of the DA-RWs increase as we increase the ion temperature while decrease with an increase in the
value of electron temperature; (b) physically, the nonlinearity of the plasma medium enhances with ion temperature
by developing a gigantic DA-RWs associated with DAWs in unstable parametric region while the nonlinearity
of the plasma medium reduces with electron temperature by developing a smaller DA-RWs associated with DAWs
in unstable parametric region. This result is a nice agreement with Rahman \textit{et al.} \cite{Rahman2018a} work.
\section{Conclusion}
\label{3sec:Conclusion}
In this study, we have performed a nonlinear analysis of DAWs in an EPIDPM  having inertial massive dust grains and
inertialess $q$-distributed electrons as well as iso-thermal positrons and ions. The nonlinear properties of the plasma
medium as well as the mechanism to generate gigantic DA-RWs associated with DAWs are governed by the standard NLSE.
It has been seen that iso-thermal ions and positrons as well as non-extensive electrons have essential influence on the MI of
DAWs. The dependency of the height and thickness of the DA-RWs on the various plasma parameters, namely, charge state of
the negatively charged dust, number density of the positron and dust grains as well as the temperature of the ion and electron
is also examined. To conclude, the results of our present investigation might be useful to understand the nonlinear
phenomena (viz., MI of DAWs and DA-RWs) in space dusty plasmas, viz., active galactic nuclei \cite{Esfandyari-Kalejahi2012,Jehan2009},
pulsar magnetospheres  \cite{Esfandyari-Kalejahi2012,Jehan2009}, interstellar
clouds \cite{Esfandyari-Kalejahi2012,Jehan2009}, supernova environments \cite{Esfandyari-Kalejahi2012,Jehan2009},
our early universe \cite{Banerjee2016}, the inner regions of the accretion disks surrounding the black hole
as well as in laboratory experiments of cluster explosions by intense laser beams \cite{Higdon2009}.

\end{document}